\documentclass[journal]{IEEEtran}

\ifCLASSINFOpdf
\else
\fi

\usepackage{color}
\usepackage{textcomp}

\usepackage{amsmath}
\usepackage{amssymb}
\usepackage{graphicx}
\usepackage{tikz}
\usepackage{multicol}
\usepackage{pifont}
\usepackage{enumerate}

\newtheorem{myproposition}{\it Proposition}

\newtheorem{mydefinition}{\it Definition}
\newtheorem{myexample}{\it Example}

\makeatletter
\def\footnoterule{\relax%
  \kern-5pt
  \hbox to \columnwidth{\hfill\vrule width 1\columnwidth height 0.8pt\hfill}
  \kern4.6pt}
\makeatother

\begin{document}
\title{Strange Attractor for Efficient Polar Code Design}

\author{{Sinan Kahraman}
\thanks{This work was supported by the Scientific and Technological Research Council of Turkey (T\"{U}B\.ITAK), grant: 1929B011500065.}
\thanks{Author is a Postdoc researcher in the Department of Electrical and Electronics Engineering, Bilkent University,
Ankara 06800, Turkey.
(e-mail: sinankahraman@gmail.com)}
}

\markboth{DRAFT}%
{DRAFT}


\maketitle

\begin{abstract}
This paper presents a definition of a construction for long polar codes.
Recently, we know that partial order is a universal property of the construction with a sublinear complexity for polar codes.
In order to describe the partial order, addition and left-swap operators are only defined as universal up to now. 
In this study, we first propose $1+\log_2 \log_2 N$ universal operators to describe multiple partial order for the block length $N=2^n$.
By using these operators, some known antichains can be universally ordered.

Furthermore, by using a simple geometric property of Gaussian approximation, we define an attractor that is a pre-defined subset of synthetic channels. They are universally less reliable than the natural channel $W$.   
Then, we show that the cardinality of this attractor is $(n+2)$-th Fibonacci number which is a significantly large number of channels for long codes.
The main contribution is that there are significant number of synthetic channels explicitly defined as almost useless by the help of attractor and multiple partial order.
As a result, proposed attractor with multiple partial order can be seen as an efficient tool to investigate and design extremely long codes.  
\end{abstract}


\IEEEpeerreviewmaketitle

\section{Introduction} 

The polar coding is the first provable coding technique to achieve the channel capacity for binary discrete memoryless channels under a quasi-linear complexity encoding, decoding and code construction methods defined in detail \cite{arikan_channel_2009}. This technique with great interest due to this important advantage has been discussed in 3GPP standardization works and accepted to be used in 5G technology. In industry, this result clearly demonstrates that the polar coding can be considered for different technologies where long codes are preferred at high transmission rates.

Briefly, the conventional construction problem for polar codes is based on the determination of the order of reliabilities for all synthetic channels. A channel-specific code construction for polar codes is described in detail by using Monte--Carlo simulation in \cite{arikan_channel_2009}. Later, for the reason that polar coding is a channel specific technique, the code construction problem has been studied for polar codes as a research topic in the literature. This is mainly due to the fact that once the polar code is designed for communication systems, it is necessary to make this design specific to the channel. For this reason, various construction methods based on calculating the reliability of the synthetic channels are discussed by density evolution \cite{mori_density_eval}, upgrading and downgrading
\cite{tal_how_2013}, and Gaussian approximation \cite{trifonov_gaussian}. A comperative study in \cite{harish} investigates the performance of these polar code constructions.

In 3GPP meetings, R1-1700088 proposed polar code design for control channel in \cite{huawei}. It is presented as a practical rule, formulated for a specific operational signal to noise ratio (SNR) region and associated transmission rates. By using this practical method, the necessity of keeping the channel indices in the hardware for the code design has become obsolete. 

Recently in \cite{schurch_partial} and \cite{barded_partial}, a partial order for synthetic channels is defined as universal (i.e. channel independent). This feature has been considered in \cite{mondelli_sublinear} to reduce the complexity of polar code design based on the considered calculations. As a result, it has been shown in \cite{mondelli_sublinear} that code design for polar codes can be done with very low complexity such as a sublinear complexity.

Here, we can briefly introduce the main contributions presented in this paper. The first contribution concerns only two operators that we know for partial order. In this work, $1+ \log_2 \log_2 N$ universal operators are defined for the block length $N=2^n$. Some synthetic channels that can not be universally ordered with partial order defined in \cite{schurch_partial} and \cite{barded_partial} can be universally ordered by our proposed operators. The number of reliability calculations introduced in \cite{mondelli_sublinear} could be reduced by exploiting the provided multiple partial order in this paper. The definition of the first attractor for polar codes is another important contribution of this study. The synthetic channels identified by this attractor are less reliable than the natural channel $W$. It is shown that the number of them are related to the Fibonacci numbers, increasing as the number of polarization steps $n$. Finally, it has been considered that an efficient design for very long codes can be made using this attractor and the mentioned multiple partial order method.

It has been shown that new operators defined by multiple partial order can compare smaller differences than the known two operators. Thus, some antichains in \cite{mondelli_sublinear} could be ordered universally and hence the number of antichains is reduced.

The considered attractor is based on a useful geometric feature obtained through the Gaussian approximation method. This feature helps to define a simple constraint for low reliable synthetic channels.
It has been shown that the subset of the synthetic channels defined by the multiple partial order feature and the attractor can be effectively used for the design of long codes. It is also considered that this advantage can be used for the non-binary polar codes in \cite{sinan}.

Paper is organized as follows, Section~\ref{sec2} introduces the system models and details for the construction of polar codes, Section~\ref{sec3} proposes new operators to define multiple partial order, Section~\ref{sec4} provides a simplified Gaussian approximation, Section~\ref{sec5} defines an attractor as a pre-defined subset of synthetic channels, Section~\ref{sec6} and~\ref{sec7} provide a discussion and some concluding remarks.

\section{System Model and Preliminary Details}\label{sec2}

The structure of the polar codes in \cite{arikan_channel_2009} with the block length $N$ is considered by $G=F^{\otimes n}$ matrix defined by the $n^{th}$ Kronecker power of $2\times 2$ kernel matrix $F$. Encoding task is expressed by $x=uG$, which is defined in modulo-2 arithmetic.
Fortunately, polar coding in \cite{arikan_channel_2009} has low complexity tasks for encoding and decoding with $\mathcal{O}(N\log N)$ complexity, taking advantage of this FFT-like structure for any discrete memoryless channels. The input vector $u$ with $N$ length has $K$ information components and $N-K$ frozen components for transmission rate $R=\frac{K}{N}$. Positions of the frozen components are known by the receiver. Here, $K$ highly reliable synthetic channels are used to carry information. $N-K$ synthetic channels with the lowest reliability are reserved for frozen. High reliability is considered as large mutual information, small Bhattacharya parameter and small error probability. The problem of code design for polar coding asks which synthetic channels should be reserved for freezing. The transition probabilities of the synthetic channels obtained after one-step of polarization are defined as follows:
\begin{eqnarray}
W^-=W(y_1,y_2|u_1)=\frac{1}{2}\sum_{u_2=0}^{1}W(y_1|u_1\oplus u_2)W(y_2|u_2),\nonumber   \\
W^+=W(y_1,y_2,u_1|u_2)=\frac{1}{2}W(y_1|u_1\oplus u_2)W(y_2|u_2),\nonumber
\end{eqnarray}
where $y_1$ and $y_2$ are noisy observation of the receiver unit. $u_1$ and $u_2$ are the input of one-step polarization. Here, $W^-$ denotes the synthesized bad channel and $W^+$ is the synthesized good channel. $W^+$ has better reliability than $W^-$ and this is expressed as $W^-\prec W \prec W^+$, where $W$ is the natural channel. Reliability ordering for $N$ synthetic channels depends on the channel. For this reason, channel specific code design is based on Monte--Carlo simulation or density evolution calculations such as Gaussian approximation. The solution to the problem of code design is sufficient to be done only once, but the fact that this solution is channel specific is a major issue in code design. Some recent research, such as partial order, has focused on this problem and they exploited relative non-channel specific reliabilities of synthetic channels. In this respect, it has been shown that the reliability of some synthetic channels can be ranked universally independent of the channel. The following definitions of two operators are provided for this purpose. To define the operators some notation can be described here. Any synthetic channel such as $(\cdots((W^+)^-)^+\cdots)^-=W^{+-+\dots -}$ obtained by n-step polarization is mapped to index in $[0,N)$ using $1$ for $+$ and $0$ for $-$ polarization step. 
E.g., $W^{--++}: W_3$ with (0011) binary index and $W^{+--+}: W_9$ with (1001) binary index. 
\newline
Let $k_i$ be the $i^{th}$ most significant bit of the binary index of $k$. 

\begin{mydefinition}[$1^{st}$ order operator]\label{firstorder}
{\it Addition}.

If $k_i=1$ and $k_j=\ell_j$ for all $j$ where $j\neq i$, then $W_\ell\preceq W_k$.\label{def1}
\end{mydefinition}

\begin{mydefinition}[$2^{nd}$ order operator]\label{secondorder}
{\it Left swap}.

If $k_i,k_{i+t}=10$ and $\ell_i,\ell_{i+t}=01$ and also $k_j=\ell_j$ for all $j$ and $t\geq1$ where $j\neq i$ and $j\neq i+t$, then $W_\ell\prec W_k$.\label{def2}
\end{mydefinition}

For more clarity, we have the results $W_{(ab0c)}\prec W_{(ab1c)}$ and $W_{(a01b)}\prec W_{(a10b)}.$
This partial order provides a sublinear complexity code design in \cite{mondelli_sublinear} and \cite{mondelli_sublinear_arxiv}.

\newpage
\section{A New Multiple Partial Order}\label{sec3}

In this section, we introduce a new method for partial order to reduce the complexity of the code design in \cite{mondelli_sublinear_arxiv}. For this purpose, we define an advanced feature of partial order by using multiple operators. It has been shown that the proposed new feature can sort synthetic channels with a small difference in reliability that can not be separated by the known partial order with the {\it Definition~\ref{firstorder}} and {\it \ref{secondorder}}. Hence the new ordering is still universal. We define new operators as follows to introduce the multiple partial order method.

\begin{mydefinition}
We assume that $E_i$ is a partition with the length of $2^{i-1}$ for $i^{th}$ order operator for $i=2,3,\dots n$ and $E_1=0$.
We define $E_{i+1}=E_{i}|E^*_{i}$ as a concatenation of $E_{i}$ and $E^*_{i}$, where $E^*_{i}$ is the binary complement of $E_{i}$. For $i=2,3,\dots n$
$$W_{E^*_{i}|E_{i}}\prec W_{E_{i}|E^*_{i}}$$
is the multiple partial order. 
\end{mydefinition}

This is a natural result of the left swap operator in {\it Definition~\ref{secondorder}}. Number of operators that can be given for the block length $N$ is $1 + \log_2 \log_2 N$. There are $5$ operators for $N=2^{16}$ are given in Table~\ref{tab_n16}. This result show that new feature has more operator than the known partial order for $N>2$.  
\begin{table}[hp]
\caption{Multiple operators for $N=2^{16}$}
\begin{center}
\begin{tabular}{|c|c|}
\hline
order & operator (less reliable $\rightarrow$ high reliable) \cr \hline
$1^{st}$ & $0\rightarrow 1$  \cr \hline
$2^{nd}$ & $01\rightarrow 10$  \cr \hline
$3^{rd}$ & $0110\rightarrow 1001$  \cr \hline
$4^{th}$ & $01101001\rightarrow 10010110$  \cr \hline
$5^{th}$ & $0110100110010110\rightarrow 1001011001101001$  \cr \hline
\end{tabular}
\end{center}
\label{tab_n16}
\end{table}%

Following examples can be given as a result of the new feature. 
\begin{myexample}\label{example3rd}
By using $3^{rd}$ order operation, 
$$W_{(0110)}\prec W_{(1001)}.$$
\end{myexample}
\begin{myexample}
By using $4^{rd}$ order operation, 
$$W_{(01101001)}\prec W_{(10010110)}.$$
\end{myexample}
They are universal partial order that can not be shown by the  {\it Definition~\ref{firstorder}} and {\it \ref{secondorder}}. Notice that the resolution of the new operators are higher than the previous definitions as follows.
$$W_{(1001)}-W_{(0110)} \prec W_{(1001)}-W_{(0101)} $$ 
This is an important property that we can exploit high resolution property to order antichans. Hence, the complexity of the code design can be reduced by using new multiple partial order method.   
  
\begin{myexample}
$\{W_{(0110)}, W_{(1001)}\}$ is an antichain for the partial order in \cite{mondelli_sublinear}. It can be universally ordered as given in {\it Example~\ref{example3rd}} by the help of higher order operation.  
\end{myexample}

Furthermore, conditional ordering can be considered as a useful method to reduce the complexity of the code design in this way. 
\subsection{Conditional Ordering}
For this purpose, the idea is based on the antichans that must be calculated for a given channel conditions to order them.  
Suppose that they are antichains that we can not order by using multiple partial order method given here. 
$$\{W_{(abcd)}, W_{(efgh)}\}\textrm{ and }\{W_{(xvyz)}, W_{(efgh)}\}$$
Suppose that the calculation of the antichain provide us a $W_{(abcd)}\prec W_{(efgh)}$ sorting and hence we can directly give a sort for the other antichain $W_{(xvyz)}\prec W_{(efgh)}$ without using any calculation. This method can be seen as a way to reduce the channel specific characteristic of the code design. For a given operational region it can be efficiently used to reduce the complexity of the code design. The following discussion is on the conditional ordering for a given operational region.       

\subsection{Discussion on the polar code design for control channel}
\begin{figure}[t!]
\centering
\includegraphics[width=0.5\textwidth]{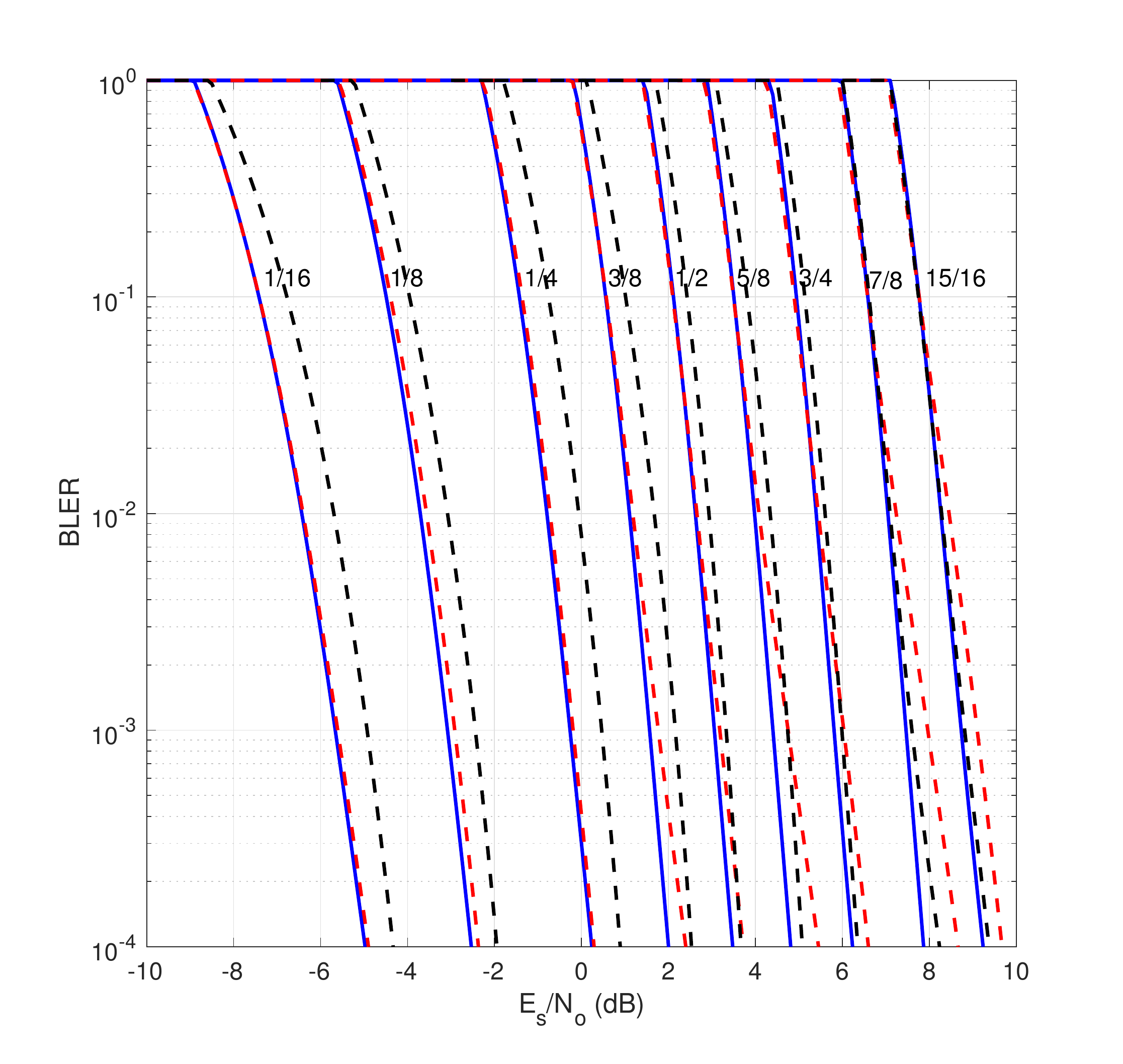}
\caption{Upper bounds for the comparison of design methods: (blue line) for Gaussian approximation and (red dashed line) for design rule in \cite{huawei} and (black dashed line) for the new design rule by using $2^{1/5\cdot i}$. } 
\label{fig_hua}
\end{figure}

In this subsection, we consider design rule in \cite{huawei}. 
As a practical method, this is based on the second order operator for the partial order and the idea of conditional ordering by using a factor (1/4) for the control channel. 
In Fig.~\ref{fig_hua}, it can be noticed that there is a significant performance loss for a specific case. 
In this way, code design rule is based on the ranking formula $\sum_{i=1}^n k_i 2^{1/4i}.$ for synthetic channels that is not a channel specific method. For this purpose, synthetic channel indices can be ordered by using this formula in \cite{huawei}. 
We focus on the particular characteristics of the selected frozen indices by the code design rule in \cite{huawei}. Here, transmission rate $R=\frac{15}{16}$ is considered that the case has a significant performance loss in Fig.~\ref{fig_hua}. 

As an observation, we notice that the number of 1 in binary expansion for the selected frozen indices has different characteristics for the considered construction rule in \cite{huawei} and Gaussian approximation for density evolution in \cite{trifonov_gaussian}.

\begin{enumerate}[i)]
\item Construction \cite{huawei}: there are $10,29,23,1$  frozen indices with the number of 1 in binary expansion = $1,2,3,4$.
\item Gaussian approx.: there are $10,36,17,0$  frozen indices with the number of 1 in binary expansion = $1,2,3,4$.  
\end{enumerate}

This can be considered as the construction rule in \cite{huawei} should increase the effect of the 1 bits in binary expansion of the indices. For this purpose we modify the construction rule by using $2^{1/5\cdot i}$ instead of $2^{1/4\cdot i}$. 

\begin{enumerate}
\item[iii)] The new design: there are $10,34,19,0$  frozen indices with the number of 1 in binary expansion = $1,2,3,4$.
\end{enumerate}

Hence, we provide the following results for the new construction (black dashed line) that are significantly better for the high rates. The new design is worse than the original method for lower rates. This means that the considered design methods are optimized for a specific case such as control channel scenario for 5G standardization. In reality, the design rule in \cite{huawei} is still a channel specific way for construction of polar codes. 

Investigation shows that it is good for the control channel but it should be re-considered for different scenarios by tuning the factor (1/4). 

\section{A Simplified Gaussian Approximation}\label{sec4}

The calculation of the reliability of each synthetic channel to design polar codes is a widely known deterministic method. For this purpose, in \cite{trifonov_gaussian} it was shown that the reliability of the synthetic channels can be efficiently calculated by using Gaussian appoximation for polar codes. This method is summarized as follows.

The Gaussian approximation for density evolution was first proposed by Chung \textsl{et al.} for low density parity check codes in \cite{Chung}. Then, the description of Gaussian approximation algorithm is given for polar code construction in \cite{trifonov_gaussian} as follows. We first assume all-zero codeword transmitted to the receiver. The log-likelihood ratio (LLR) for a noisy observation $y_i=x_i+n_i$ is defined as $L^i_1(y_i)=\log \frac{W(y_i|0)}{W(y_i|1)}$.    
The probability density function is $f(x)=e^{-x^2/2\sigma^2}$ for additive white Gaussian noise with $N(0,\sigma^2)$ distribution. Now, it can be considered that the expected value of the likelihood ratio $E\left[L^i_1(y_i)\right]$ as follows.
\begin{eqnarray}
E\left[L^i_1(y_i)\right]&=&E\left[\log \frac{W(y_i|0)}{W(y_i|1)}\right],\nonumber\\
&=&E\left[\log \frac{e^{-{(y_i-1)^2}/2\sigma^2}}{e^{-{(y_i+1)^2}/2\sigma^2}}\right],\nonumber\\
&=&E\left[\log \frac{e^{{(-y_i^2+2y_i-1)}/2\sigma^2}}{e^{{(-y_i^2-2y_i-1)}/2\sigma^2}}\right],\nonumber\\
&=&E\left[\log {e^{{4y_i}/2\sigma^2}}\right],\nonumber\\
&=&E\left[{{4y_i}/2\sigma^2}\right],\nonumber\\
&=&{{4E\left[y_i\right]}/2\sigma^2}.\nonumber
\end{eqnarray}
Finally,
$$E\left[L^i_1(y_i)\right]={{2}/\sigma^2}.$$

Variance of the likelihood ratio is given as 
\begin{eqnarray}
V\left[L^i_1(y_i)\right]&=&E\left[\left(L^i_1(y_i)-\frac{2}{\sigma^2}\right)^2\right],\nonumber\\
&=&E\left[\left(\frac{2(y_i-1)}{\sigma^2}\right)^2\right]\nonumber,\\
&=&\frac{4\sigma^2}{\sigma^4}.\nonumber
\end{eqnarray}
Finally,
$$V\left[L^i_1(y_i)\right]=\frac{4}{\sigma^2}.$$

The update rules for the expectations of inter-level LLRs is given for $i=1,\dots,n/2$ as follows 
\begin{eqnarray}
E\left[L^{(2i-1)}_{j}\right]&=&\phi^{-1}\left(1-\left(1-\phi\left(E\left[L^i_{j/2}\right]\right)\right)^2\right),\nonumber \\
E\left[L^{(2i)}_{j}\right]&=&2E\left[L^i_{j/2}\right]\nonumber
\end{eqnarray}
where
$$\phi(x)=\Bigg \{
  \begin{tabular}{lr}
  $1-\frac{1}{\sqrt{4 \pi x}} \int_{-\infty}^{\infty} \tanh \frac{u}{2} e^{-\frac{(u-x)^2}{4x}} du$ & $x>0$ \\
  $1$, & $x=0$  
  \end{tabular}.$$
  
The error probability of indices $i\in\{1,\dots,N\}$ is given as follows 
$$\pi_i \approx Q\left(\sqrt{E\left[L^i_{N}\right]/2}\right)=\frac{1}{2}\textrm{erfc}\left(\frac{1}{2}\sqrt{E\left[L^i_{N}\right]}\right)$$
where $$\textrm{erfc}\left(x\right)=\frac{2}{\sqrt{\pi}} \int_{x}^{\infty} e^{-v^2} dv.$$
An upper bound of the error probability is the sum of error probabilities for the set of information indices.

To simplify the update rule we use an approximation of
$$\tanh x \approx \left\{
  \begin{tabular}{cr}
  $1$, & $x>0$ \\
  $0$, & $x=0$ \\ 
  $-1$, & $x<0$
  \end{tabular}\right .$$
as given in Appendix I, and hence, the simplified update rule can be provided by using the following definitions.
\begin{eqnarray*}
\phi(x) =\textrm{erfc}\left(\frac{\sqrt{x}}{2}\right)
\end{eqnarray*}
\begin{eqnarray*}
{\phi}^{-1}(x) = 4 \left(\textrm{erfcinv}\left(x\right)\right)^2
\end{eqnarray*}
The simplified update rule is given as follows:
\begin{eqnarray}
&&E\left[L^{(2i-1)}_{j}\right]=\nonumber\\
&&4\left(\textrm{erfcinv}\left(1-\left(1-\textrm{erfc}\left(\frac{1}{2}\sqrt{E\left[L^i_{j/2}\right]}\right)\right)^2\right)\right)^2,\nonumber
\end{eqnarray}
\begin{eqnarray}
&&E\left[L^{(2i)}_{j}\right]=2E\left[L^i_{j/2}\right].\nonumber
\end{eqnarray}

This is numerically stable method. Additionally, this can be efficiently implemented by using a lookup table for the function $\textrm{erfc(x)}$ and $\textrm{erfcinv(x)}$.   
In Appendix I, we verify that the simplified method is accurately close to the original Gaussian approximation (Chung's) method. Furthermore, these are close to the simulation results for additive white Gaussian channel.

\section{A Strange Attractor}\label{sec5}
\begin{figure}[hp]
\centering
\includegraphics[width=0.5\textwidth]{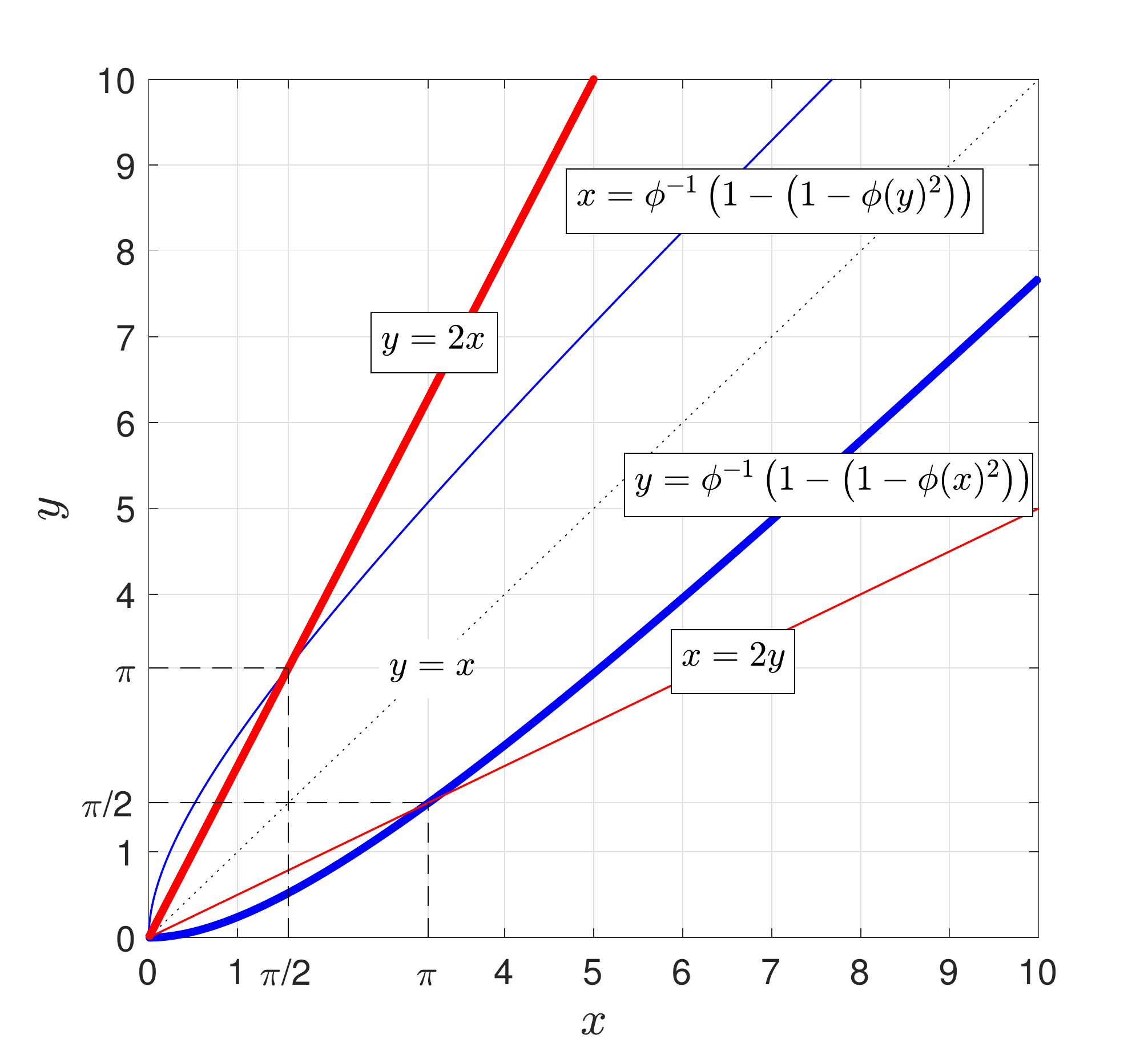}
\caption{Plot of the recursive functions for update rule of Gaussian approximation method. (bold curves: the functions and thin curves: the reflections.)} 
\label{fig1}
\end{figure}

In the previous section, the reliability of synthetic channels was precisely computed by using the simplified Gaussian approximation method. In this section, we investigate the geometry of the functions that define the recursive update rule in the Gaussian approximation method to understand the universal properties of synthetic channels.
First, $y=2x$ and $y=\phi^{-1}\left(1-\left(1-\phi\left(x\right)\right)^2\right)$ functions are depicted in Fig.~\ref{fig1} and the reflections of these curves according to $y=x$ line are also added.
\newline
Here, we present some observations about the geometric properties of functions as follows. Let functions be defined as $f_1(x)=x/2$ and  $f_2(x)=\phi^{-1}\left(1-\left(1-\phi\left(x\right)\right)^2\right)$. 
\begin{enumerate}[i)]
\item $y=f_1(x)$ and $y=f_2(x)$ intersect at $(x=0,y=0)$.
\item $y=f_1(x)$ and $y=f_2(x)$ intersect at $(x=\pi,y=\pi/2)$.
\item $f_1(x)>f_2(x)$ for $x \in (0,\pi)$.
\end{enumerate}

As a result of these observations, we can identify synthetic channels that are worse than the natural channel using a simple constraint with an attractor.

\begin{mydefinition}
Attractor is a subset of pre-defined synthetic channels that are universally unreliable than the natural channel $W_k\prec W$. 
\end{mydefinition}
Any binary discrete memoryless channel with a limited LLR can be polarized to the synthetic channels with the index without 11 partition in binary expansion defines an attractor.

\begin{myexample}
Assume that the natural channel $W$ has $LLR<\pi/2$. The synthetic channels with the index that the binary representation without 11 partition are universally unreliable than the natural channel.
\end{myexample}
\begin{myexample}\label{example_pi}
Assume that the natural channel $W$ has $LLR<\pi$. The synthetic channels with the index that the binary representation without 11 partition and the first bit is also not 1 are universally unreliable than the natural channel.
\end{myexample}

\begin{myproposition}\label{th_attractor}
As the block length increases, the LLR value of synthetic channels that do not have 11 partition converges to 0 (i.e., unreliable) and the number of these synthetic channels for $N=2^n$ block length is $F_{2+n}$, where $F_i$ is $i^{th}$  Fibonacci number in $\{1,1,2,3,5,8,13,21,34,55,89,\dots\}$.
\end{myproposition}

The proof for the proposition is presented in two parts. The first part of the proof is concerned with the exact number of n-length bit strings that do not have 11 partitions. Let $\mathcal{A}^i$ be a set of $i$-long bit strings that do not have 11 partitions. The exact number of the strings $|\mathcal{A}^i|$ can be given as follows:

\begin{enumerate}[i)]
\item $|\mathcal{A}^1|=2$ where $\mathcal{A}^1:\{0,1\}$
\item $|\mathcal{A}^2|=3$ where $\mathcal{A}^2:\{00,01,10\}$
\item $|\mathcal{A}^3|=5$ where $\mathcal{A}^3:\{000,001,010,100,101\}$
\item $|\mathcal{A}^4|=8$ where\\ $\mathcal{A}^4:\{{\bf 0}000,{\bf 0}001,{\bf 0}010,{\bf 0}100,{\bf 0}101,{\bf 10}00,{\bf 10}01,{\bf 10}10\}$\\\vdots
\item $|\mathcal{A}^\ell|=|[{\bf 0}|\mathcal{A}^{\ell-1}],[{\bf 10}|\mathcal{A}^{\ell-2}]|=|\mathcal{A}^{\ell-1}|+|\mathcal{A}^{\ell-2}|$ (Fibonacci). 
\end{enumerate}
As a result, $|\mathcal{A}^n|=F_{n+2}$ 

where $F_n=\{1,1,2,3,5,8,13,21,34,55,89,144,\dots\}$.

The second part of this proof is about the attractor. As a result of the observations, we could pre-define $F_{n+2}$ synthetic channels named as attractor thanks to the geometrical properties in Fig.~\ref{fig1} we obtained. It is clear to see that LLR goes to 0 for all possible bit strings that do not have 11 partitions. The geometric interpretation of this result is given by the following figure.

\begin{figure}[hp]
\centering
\includegraphics[width=0.5\textwidth]{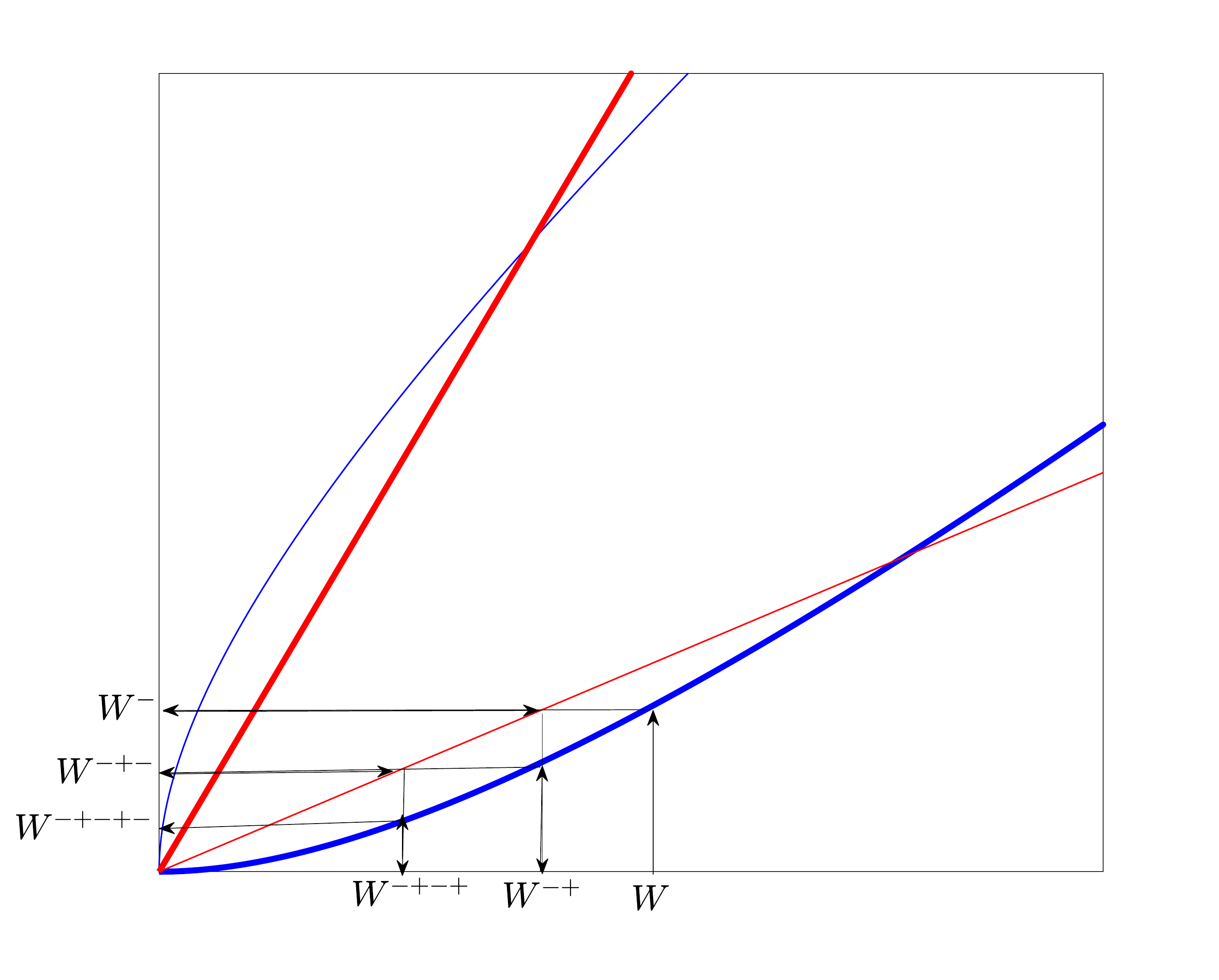}
\caption{Example of synthetic channels that are worse than natural channel.} 
\label{fig12}
\end{figure}

The number of such a bad synthetic channels are provided for a given $n$ the number of polarization steps in Table~\ref{tab1}.

Now let's examine the asymptotic behaviour of the number of these bad channels, which we are pre-defined here by using the attractor. For this purpose, we provide the following expression. 

$$\lim_{N\rightarrow \infty}{\frac{\textsl{Number of channels with (11)}}{\textsl{Number of all channels}}}=1.$$

Proof is given here. We consider the expression as follows.
The exact number channels with 11 partition can be described as follows:
$$\Delta=\Delta_1+\Delta_2$$
where $\Delta_1$ is shown in Fig.~\ref{fig_delta1} and $\Delta_2$ is shown in Fig.~\ref{fig_delta2}. 

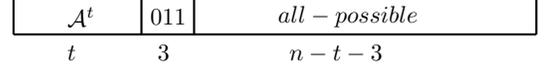
\begin{figure}[hp]
\centering
\vspace{0.5cm}
\begin{tikzpicture}[thick,scale=0.5, every node/.style={scale=0.9}]
\draw (-0.2,-0.5)  -- (13.7,-0.5);
\draw (-0.2,0.5)  -- (13.7,0.5);
\draw (-0.2,-0.5)  -- (-0.2,0.5);
\draw (3.2,-0.5)  -- (3.2,0.5);
\draw (4.6,-0.5)  -- (4.6,0.5);
\draw (13.7,-0.5)  -- (13.7,0.5);
\draw (3.2,0-0.03) node[right] {$011$};
\draw (3.4,-1) node[right] {$3$};
\draw (6.6,0-0.03) node[right] {$all-possible$};
\draw (6.9,-1) node[right] {$n-t-3$};
\draw (1,0-0.03) node[right] {$\mathcal{A}^t$};
\draw (1,-1) node[right] {$t$};
\end{tikzpicture}
\caption{A graphical representation of the case $\Delta_1$.}
\label{fig_delta1}
\end{figure}
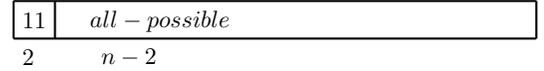
\begin{figure}[hp]
\centering
\vspace{0.5cm}
\begin{tikzpicture}[thick,scale=0.5, every node/.style={scale=0.9}]
\draw (-0.2,-0.5)  -- (13.7,-0.5);
\draw (-0.2,0.5)  -- (13.7,0.5);
\draw (-0.2,-0.5)  -- (-0.2,0.5);
\draw (0.9,-0.5)  -- (0.9,0.5);
\draw (13.7,-0.5)  -- (13.7,0.5);
\draw (-0.2,0-0.03) node[right] {$11$};
\draw (-0.2,-1) node[right] {$2$};
\draw (1.6,0-0.03) node[right] {$all-possible$};
\draw (1.9,-1) node[right] {$n-2$};
\end{tikzpicture}
\caption{A graphical representation of the case $\Delta_2$.}
\label{fig_delta2}
\end{figure}
Here,
$$\Delta_1=\sum_{t=0}^{n-3} |\mathcal{A}^t| \cdot 2^{N-t-3}$$
and 
$$\Delta_2=2^{n-2}.$$

We can show that
$$\Delta=2^{n-2}+\sum_{t=0}^{n-3} |\mathcal{A}^t| \cdot 2^{n-t-3}$$

\begin{eqnarray}
\Delta&=&\left[\sum_{t=0}^{n-3} F_{t+2} \cdot 2^{n-t-3}\right]+2^{n-2}\nonumber \\
&=&2^{n-1}\left(\left[\sum_{t=0}^{n-3} F_{t+2}/2^{t+2}\right]+1/2\right)\nonumber \\
&=&\left[\sum_{t=0}^{n-3} F_{t+2} \cdot 2^{n-t-3}\right]+2^{n-2}\nonumber \\
&=&2^{n-1}\left(\left[\sum_{t=0}^{n-3} F_{t+2}/2^{t+2}\right]+1/2\right).\nonumber 
\end{eqnarray}
Then, we have
$$\Delta=2^{n-1}\left(\left[\frac{F_2}{2^2}+\frac{F_3}{2^3}+\dots+\frac{F_{n-1}}{2^{n-1}}\right]+\frac{F_1}{2^1}+\frac{F_0}{2^0}\right).$$

Here, notice that $\frac{F_0}{2^0}=0$ and $\frac{F_1}{2^1}=1/2$.
The final exact expression is
$$\Delta=2^{n-1}\sum_{t=0}^{n-1}F_t/2^t.$$
There is power serie $\sum_{t=0}^{\infty} F_{t} \cdot k^{-t}=\frac{k}{k^2-k-1}$ for integer $k>1$.

As a result, 
$$\lim_{n\rightarrow \infty}{\frac{\textsl{Number of channels with (11)}}{\textsl{Number of all channels}}}=\lim_{n\rightarrow \infty} \frac{2^n\frac{1}{2}\frac{2}{2^2-2-1}}{2^n}=1.$$

Now we can consider here that we can benefit from the definition of attractor in the design of long polar codes. In this direction, we can separate the problem into two parts. 

The first part discloses a situation in which the natural channel LLR parameter is greater than $\pi/2$.
In this case, the indexes of the synthetic channels to be identified by the attractor will start from the most significant bit position, and the different length sequences will be determined which will reduce the LLR value of the natural channel to less than $\pi/2$ as an inter-level LLR value, (please see the {\it Example~\ref{example_pi}}).

As a second step, n-length binary expansions are obtained with complementary arrays with no 11 partitions for the specified arrays.
Thus, it is stated that the attractor can be defined for different LLR values that the natural channel has.

These are not all of them but they are significant number of synthetic channels. They are universally worse than the natural channel. For more clarity, we provide the following definition of efficient design and its small example.
\begin{mydefinition}
As a plain text an efficient code design can be given as follows.
\begin{enumerate}[i)]
\item Define Attractor for $n$ as $\Omega$ is a subset of $\{1,2,\dots,N\}$
\item {\bf for} i=1,...,n\newline \,\,\,\,Apply $i$-th order operator to update $\Omega$\newline {\bf end}
\item Apply simplified Gaussian approximation for the complement of the set $\Omega$
\end{enumerate}
\end{mydefinition}

\begin{table}[t!]
\caption{Number of synthetic bad channels}
\begin{center}
\begin{tabular}{|c|c|c|}
\hline
n & Number of synthetic bad channels & Rate \cr \hline
6 & 21 & 0.6719 \cr \hline
7 & 34 & 0.7344 \cr \hline
8 & 55 & 0.7852 \cr \hline
9 & 89 & 0.8262 \cr \hline
10 & 144 & 0.8594 \cr \hline
11 & 233 & 0.8862 \cr \hline
12 & 377 & 0.9080 \cr \hline
13 & 610 & 0.9255 \cr \hline
14 & 987 & 0.9398 \cr \hline
15 & 1597 & 0.9513 \cr \hline
16 & 2584 & 0.9606 \cr \hline
\end{tabular}
\end{center}
\label{tab1}
\end{table}%

\begin{myexample}
We consider $n=6$ in this example. There are $64$ synthetic channels placed in the following table. Here, black bold face binary expansion denotes the bad channel associated by the attractor (i.e. they do not have 11). There are $F_{n+2}=21$ bad channels for $n=6$ that are worse than the natural channel.

Now, we can apply the multiple partial order to find more channels that are worse than the natural channel $W$.
When we consider the first order operator to increase the number of bad channels, it can be noticed that there are not any new bad channel by removing 1 in the attractor. The result is guaranteed that it is placed in the attractor. Then, we can apply second order operator to find more synthetic channels.

For example; $W_{(101000)}$ is a member of the attractor (i.e., 01$\rightarrow$10). By using 2nd order operator, we have the following result. 
$$W_{(011000)}\prec W_{(101000)}\prec W.$$

Finally, we can apply third order operator.

For example; $W_{(011100)}$ is a member of the bad channels that are union set of attractor and 2nd order operator. By using 3rd order operator (i.e., 0110$\rightarrow$1001), we have the following result. 
$$W_{(011100)}\prec W_{(101010)}\prec W.$$
The synthetic channels found by multiple partial order are denoted by blue bold face in Table.~\ref{tab2}.

As a result, we have found $35$ synthetic channels in $64$ that are worse than the natural channels for $LLR<\pi/2$. On the other hand $25$ of them is still worse than the natural channel for $LLR<\pi$ (i.e., they do not have 1 in the first bit).  
\end{myexample}

\begin{table}[hp]
\caption{Example for Attractor and Multiple Partial Order}
\begin{center}
\begin{tabular}{|c|c|c|c|}
\hline
                  \bf 000000 &                   \bf 001000 &                  \bf 010000 & \color{blue} \bf 011000 \cr \hline
                  \bf 000001 &                   \bf 001001 &                  \bf 010001 & \color{blue} \bf 011001 \cr \hline
                  \bf 000010 &                   \bf 001010 &                  \bf 010010 & \color{blue} \bf 011010 \cr \hline
\color{blue}\bf 000011 & \color{gray}      001011 &\color{blue} \bf 010011 &\color{gray}      011011 \cr \hline
                   \bf 000100 & \color{blue} \bf 001100 &                 \bf 010100 & \color{blue} \bf 011100 \cr \hline
                   \bf 000101 & \color{blue} \bf 001101 &                 \bf 010101 &\color{gray}      011101 \cr \hline
\color{blue} \bf 000110 & \color{blue}   \bf 001110 & \color{blue} \bf 010110 &\color{gray}      011110 \cr \hline
  \color{blue} \bf 000111 & \color{gray}      001111 &\color{gray}      010111 &\color{gray}      011111 \cr \hline \hline
                   \bf 100000 &                   \bf 101000 &\color{gray}      110000 &\color{gray}      111000 \cr \hline
                   \bf 100001 &                   \bf 101001 &\color{gray}      110001 &\color{gray}      111001 \cr \hline
                   \bf 100010 &                   \bf 101010 &\color{gray}      110010 &\color{gray}      111010 \cr \hline
\color{blue} \bf 100011 &\color{gray}      101011 &\color{gray}      110011 &\color{gray}      111011 \cr \hline
                   \bf 100100 &\color{gray}      101100 &\color{gray}      110100 &\color{gray}      111100 \cr \hline
                   \bf 100101 &\color{gray}      101101 &\color{gray}      110101 &\color{gray}      111101 \cr \hline
\color{blue} \bf 100110 &\color{gray}      101110 &\color{gray}      110110 &\color{gray}      111110 \cr \hline
\color{gray}      100111 &\color{gray}      101111 &\color{gray}      110111 &\color{gray}      111111 \cr \hline
\end{tabular}
\end{center}
\label{tab2}
\end{table}%

\section{Discussion on the Main Results}\label{sec6}
In this section we summarize the results of our work. In addition, we discuss the implications of the results and their potential use in future studies.

The first result we have is related to the universal partial order. By taking advantage of the new multi-operators we have proposed, we have been able to define high-resolution universal comparisons between synthetic channels. This result shows that anti-chain number can be reduced. Thus, the calculation complexity of the proposed code design can be reduced by the known partial order method.

The second result is related to the calculation of the reliability of synthetic channels. In this direction, a simplified method of numerically stable Gaussian approximation technique is proposed. It has also been shown that the performance of this method is very close to the known method. Thus, an efficient Gaussian approximation method has been achieved which can be used for reliability calculation in code design.

The third result we have is related to the attractor we have identified in synthetic channels. In this direction, the geometric properties of the functions defining the update rules of the Gaussian approximation method are exploited. As a result, a considerable number of bad synthetic channels can be identified with a simple constraint. This contribution can be used to reduce complexity in the design of long codes.

We can evaluate that the results obtained can be used together. The work in this direction aims to reduce the complexity of the code design problem as a framework. It can be considered that the methods proposed on the problem of designing long polar codes in particular can be used effectively.
As the future works, to design non-binary polar codes with the equidistant polarizing transforms in \cite{sinan} the strange attractor method can be used. It can also be considered for polar codes with arbitrary binary linear kernels in \cite{huang}.


\section{Conclusion}\label{sec7}
The universal features that polar codes have are an important advantage for efficient code design problems.
In this work we have defined new universal features. These features allowed higher resolution sorting.
We have made it easier to calculate the reliability of synthetic channels that are important for efficient code design. For this purpose, we simplified the recursive update functions of the Gaussian approximation method. We showed that the result of the simplified method is quite close to the original way.
We have identified an attractor for bad synthetic channels. We have pre-defined a significant number of bad synthetic channels associated with Fibonacci numbers.mk
As a result, all of these contributions we present as a framework study can be considered for efficient code design for long polar codes.

\appendices
\section{Simplification of the functions: $\phi(x)$ and $\phi^{-1}(x)$}
$$\phi(x)=\Bigg \{
  \begin{tabular}{lr}
  $1-\frac{1}{\sqrt{4 \pi x}} \int_{-\infty}^{\infty} \tanh \frac{u}{2} e^{-\frac{(u-x)^2}{4x}} du$ & $x>0$ \\
  $1$, & $x=0$  
  \end{tabular}$$
First, we consider the following assumption:
$$\tanh x \approx \left\{
  \begin{tabular}{cr}
  $1$, & $x>0$ \\
  $0$, & $x=0$ \\ 
  $-1$, & $x<0$
  \end{tabular}\right .$$
Then, we use the following equations.
\begin{eqnarray}
&&\frac{1}{\sqrt{4 \pi x}} \int_{-\infty}^{\infty} \tanh \frac{u}{2} e^{-\frac{(u-x)^2}{4x}} du \approx \nonumber\\
&&\frac{1}{\sqrt{4 \pi x}} \left( \int_{0}^{\infty} e^{-\frac{(u-x)^2}{4x}} du - \int_{-\infty}^{0} e^{-\frac{(u-x)^2}{4x}} du\right) \nonumber
\end{eqnarray}
We apply the transformation: $\frac{u-x}{2\sqrt{x}}=v$. Then, 
\begin{eqnarray}
&&\frac{1}{\sqrt{4 \pi x}} \left( \int_{0}^{\infty} e^{-\frac{(u-x)^2}{4x}} du - \int_{-\infty}^{0} e^{-\frac{(u-x)^2}{4x}} du\right)=\nonumber\\
&&\frac{1}{\sqrt{\pi}} \left( \int_{-\sqrt{x}/2}^{\infty} e^{-v^2} dv - \int_{-\infty}^{-\sqrt{x}/2} e^{-v^2} dv\right).\nonumber
\end{eqnarray}
Then, we use the definition: $$\textrm{erfc}\left(\frac{\sqrt{x}}{2}\right)=\frac{2}{\sqrt{\pi}} \int_{\sqrt{x}/2}^{\infty} e^{-v^2} dv.$$  
\begin{eqnarray}
&&1-\textrm{erfc}\left(\frac{\sqrt{x}}{2}\right)=\nonumber\\
&&\frac{1}{\sqrt{4 \pi x}} \left( \int_{0}^{\infty} e^{-\frac{(u-x)^2}{4x}} du - \int_{-\infty}^{0} e^{-\frac{(u-x)^2}{4x}} du\right).\nonumber
\end{eqnarray}
Finally, we have the simplified equations as follows:
$$\phi\left(x\right) = \textrm{erfc}\left(\frac{\sqrt{x}}{2}\right),$$

$${\phi}^{-1}\left(x\right) = 4 \left(\textrm{erfcinv}\left(x\right)\right)^2.$$ 

\begin{figure}[t!]
\centering
\includegraphics[width=0.48\textwidth]{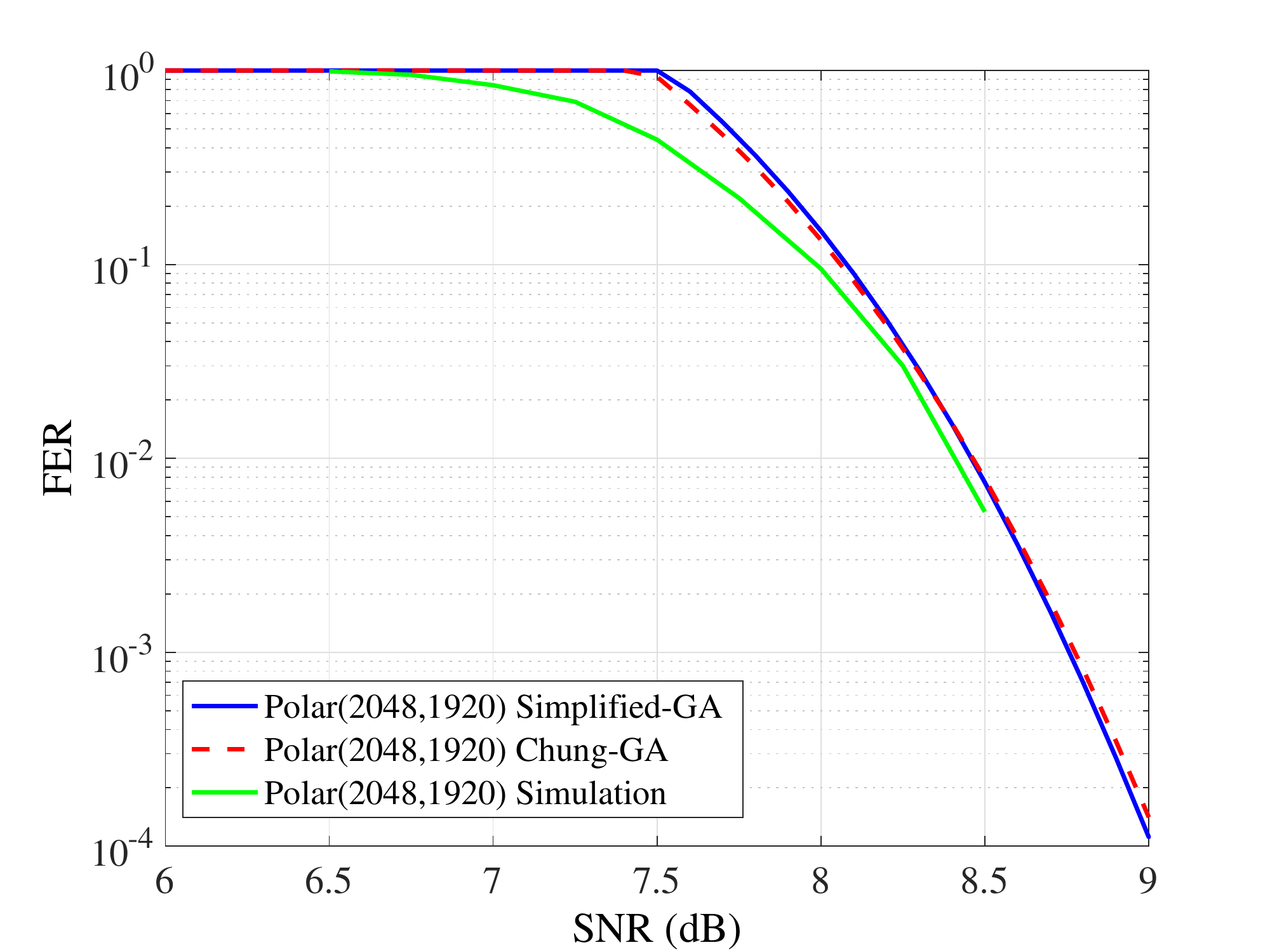}%
\caption{Simulation results and upper bounds by the Gaussian approximation (Chung's) method and the simplified-Gaussian approximation method.}
\label{fig_simpGA}
\end{figure}

\section*{Acknowledgment}
This work was supported by the Scientific and Technological Research Council of Turkey (T\"{U}B\.ITAK), grant: 1929B011500065. I would like to thank Prof. Erdal Ar\i kan and Dr. Zhiliang Huang for helpful discussions. 

\ifCLASSOPTIONcaptionsoff
  \newpage
\fi

\bibliography{attractor_Ref}{}
\bibliographystyle{IEEEtran}

\end{document}